\documentclass[twocolumn,prl,superscriptaddress,showpacs]{revtex4}
\newcommand{\pagenumbaa}{1}
\usepackage{graphicx}
\usepackage{color}

\usepackage{units}
\begin{document}


\title{Intrinsic Metastabilities in the Charge Configuration of a Double Quantum Dot}

\author{D. E. F. Biesinger}\altaffiliation{\vspace{-2mm}equally contributing authors}
\affiliation{Department of Physics, University of Basel, Klingelbergstrasse 82, CH-4056 Basel, Switzerland}

\author{C. P. Scheller}\altaffiliation{\vspace{-2mm}equally contributing authors}
\affiliation{Department of Physics, University of Basel, Klingelbergstrasse 82, CH-4056 Basel, Switzerland}
\affiliation{Swiss Federal Laboratories for Materials Science and Technology, EMPA, \"Uberlandstrasse 129, CH-8600 D\"ubendorf, Switzerland}

\author{B. Braunecker}
\affiliation{Scottish Universities Physics Alliance, University of St Andrews, North Haugh, St Andrews KY16 9SS, UK}

\author{J. Zimmerman}
\affiliation{Materials Department, University of California, Santa Barbara, CA93106, USA}

\author{A. C. Gossard}
\affiliation{Materials Department, University of California, Santa Barbara, CA93106, USA}

\author{D. M. Zumb\"uhl}
\email{dominik.zumbuhl@unibas.ch}
\affiliation{Department of Physics, University of Basel, Klingelbergstrasse 82, CH-4056 Basel, Switzerland}

\begin{abstract}
We report a thermally activated metastability in a GaAs double quantum dot exhibiting real-time charge switching in diamond shaped regions of the charge stability diagram. Accidental charge traps and sensor back action are excluded as the origin of the switching. We present an extension of the canonical double dot theory based on an intrinsic, thermal electron exchange process through the reservoirs, giving excellent agreement with the experiment. The electron spin is randomized by the exchange process, thus facilitating fast, gate-controlled spin initialization. At the same time, this process sets an intrinsic upper limit to the spin relaxation time.
\end{abstract}
\maketitle

\setcounter{page}{\pagenumbaa}
\thispagestyle{plain}



Spins in quantum dots \cite{Loss1} are promising candidates for the realization of qubits -- the elementary units of a quantum computer. Great progress was made in recent years towards implementing quantum information processing schemes with electron spins in GaAs quantum dots \cite{rev1,Petta:2005uq,Nowack:2007du,Foletti:2009hk,Brunner2011,Shulman2012,Medford2013}, which hold the potential for scaling to a large number of qubits \cite{GaudreauNP,Trifunovic,Kloeffel}. Stable qubits with long coherence times are of crucial importance to execute numerous coherent quantum gates. Spin echo and dynamical decoupling techniques were successfully employed to isolate the electronic system from the slowly fluctuating nuclear spins of the GaAs host material \cite{Petta:2005uq,Koppens:2008dx,Bluhm:2010fj,Medford2012,Dial2013}, enhancing the coherence time $T_2$ from below 1~$\mathrm{\mu s}$ to much longer times exceeding 0.2\,$\mathrm{ms}$. A fundamental limit $T_2\leq 2\,T_1$ is set by the spin relaxation time $T_1$. In a magnetic field, spins relax through spin-phonon coupling mediated by the spin-orbit interaction \cite{rev1,Golovach:2004eta,ScarlinoPRL,KornichPRB}. Since here the spin-orbit coupling is weak, very long $T_1$ times result, exceeding 1~s at $1\,$T \cite{AmashaPRL}, leaving ample room for further improvements of the spin qubit coherence.

In this Letter, we report the experimental observation of a thermal electron exchange process via the reservoirs of a quantum dot, setting an intrinsic upper bound to $T_1$ which can be orders of magnitude lower than the fundamental spin-phonon limit \cite{Golovach:2004eta}. The resulting metastable charge states -- appearing in the double dot (DD) in absence of interdot tunneling -- make the exchange process detectable with a charge sensor. Within a diamond shaped region, the DD switches its charge-state back and forth over time from an electron on the left dot to an electron on the right dot without direct interdot tunneling. After excluding unintentional charge traps and sensor back action, we present an extension of the orthodox DD transport theory accounting very well for the observations. The exchange process can be used for fast qubit initialization \cite{Shulman2012}. Finally, we outline ways to extend $T_1$ up to the spin-phonon limit.

\begin{figure}[hb]
\vspace{-8mm}\includegraphics[width=0.95\columnwidth]{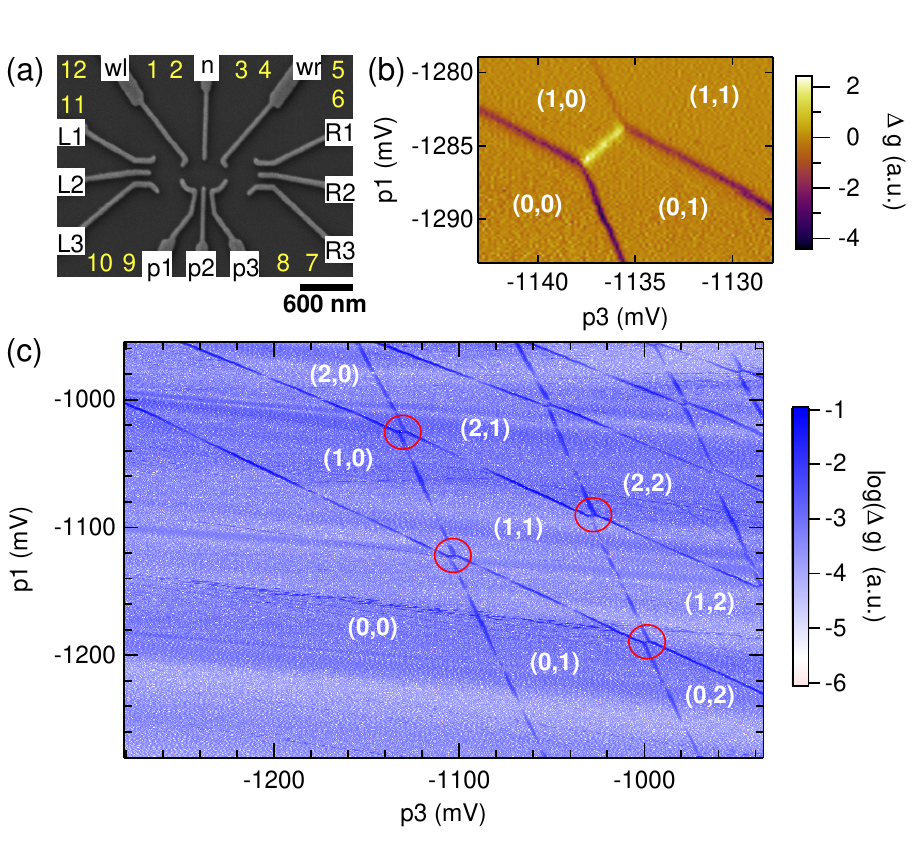}\vspace{-3mm}
\caption{(a) Scanning electron micrograph of a nominally identical device, with contacts (yellow) and gates (white boxes). Gates p1-p3, wl, wr and n are used to form the DD in the center while L1-L3 control the left and R1-R3 the right sensor dot. (b) Numerical derivative $\Delta g$ of the left sensor conductance. (c) Larger CSD with gate configuration as in Fig.\,\ref{fig:2}. Circles indicate vertices where the metastability was observed.
}
\label{fig:1}
\end{figure}

The sample is fabricated from a $GaAs$ heterostructure with a 2D electron gas 110\,nm below the surface (density $\unit[2.6\cdot 10^{11}]{cm^{-2}}$ and mobility $\unit[4\cdot 10^{5}]{cm^2/Vs}$). The device layout, see Fig.\,\ref{fig:1}(a), is adopted from Ref.\,\onlinecite{Barthel:10}. Each dot adjacent to the DD (center) acts as a charge sensor \cite{Pepper93,Elzerman:04}, changing conductance $g$ up to a factor of two upon adding one electron to the dot closer to the sensor. The high sensitivity is due to the steep edges of a Coulomb blockade peak where the sensor is biased \cite{Barthel:10}. Simultaneously, strong capacitive shifts of the sensor biasing point are seen when changing DD gate voltages. Thus, compensation with linear feedback on sensor plunger L2 is employed to maintain charge sensitivity.

The experiment was done in a dilution refrigerator at base temperature $T\sim20\,$mK. Ag-epoxy microwave filters and thermalizers \cite{Scheller2014} are mounted at the mixing chamber for improved cooling \cite{clark2010, casparis2012}, giving an electron temperature of $T_e\sim60\,$mK from Coulomb blockade thermometry \cite{Maradan14}. All data presented here was acquired with the left sensor dot, though the right sensor gives essentially the same results. The DD reservoirs are held at the same potential. The sensor is typically biased at 60\,$\mathrm{\mu V}$ DC and the resulting current is digitized in real-time with a measurement bandwidth of 10\,kHz, limited by the signal to noise ratio, not technical bandwidth.

The charge stability diagram (CSD) of the DD is shown in Fig.~\ref{fig:1}(c). Sharp lines indicate charge transitions, forming the usual DD honeycombs \cite{vanderWiel:2002}. Charge states are labeled ($N_1$, $N_2$) denoting the absolute number of electrons on the (left, right) dot in the ground state. The first two triple points are shown in Fig.\,\ref{fig:1}(b). Dark lines designate transitions involving a reservoir, while the bright line corresponds to equal left and right dot energies (zero detuning), separating the $(1,0)$ and $(0,1)$ states. The absence of curvature indicates weak interdot tunneling. These data reflect standard DD behavior as expected \cite{vanderWiel:2002}.

\begin{figure}[Ht]
\vspace{0mm}\includegraphics[width=0.95\columnwidth]{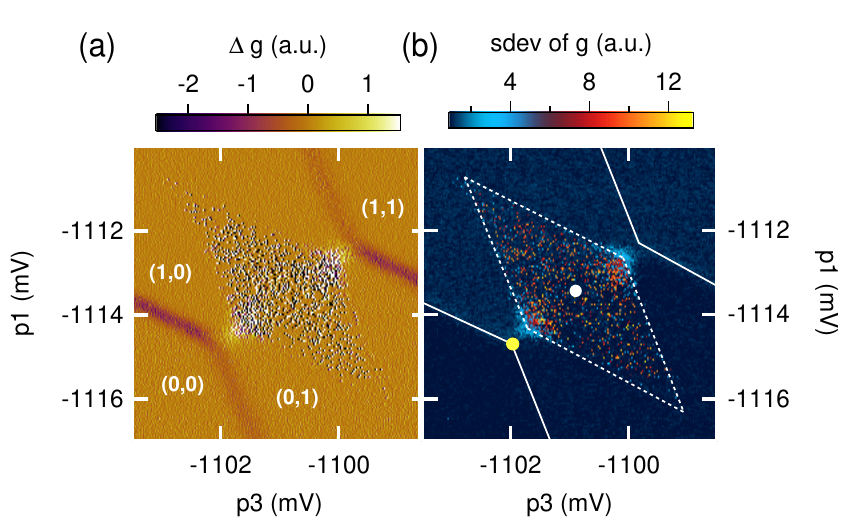}\vspace{-2mm}
\caption{Sensor $\Delta g$ (a) and standard deviation of $g$ (b) from 20\,ms measurement windows with reduced tunnel rates compared to Fig.~\ref{fig:1}(b). Dashed lines indicate the borders of the metastable diamond while solid lines show the reservoir transitions, obtained from (a). The white and yellow dots show gate voltage configurations used in Figs.~\ref{fig:3} and \ref{fig:4}.}
\label{fig:2}\vspace{-4mm}
\end{figure}

\begin{figure}[Ht]
\vspace{0mm}\includegraphics[width=0.95\columnwidth]{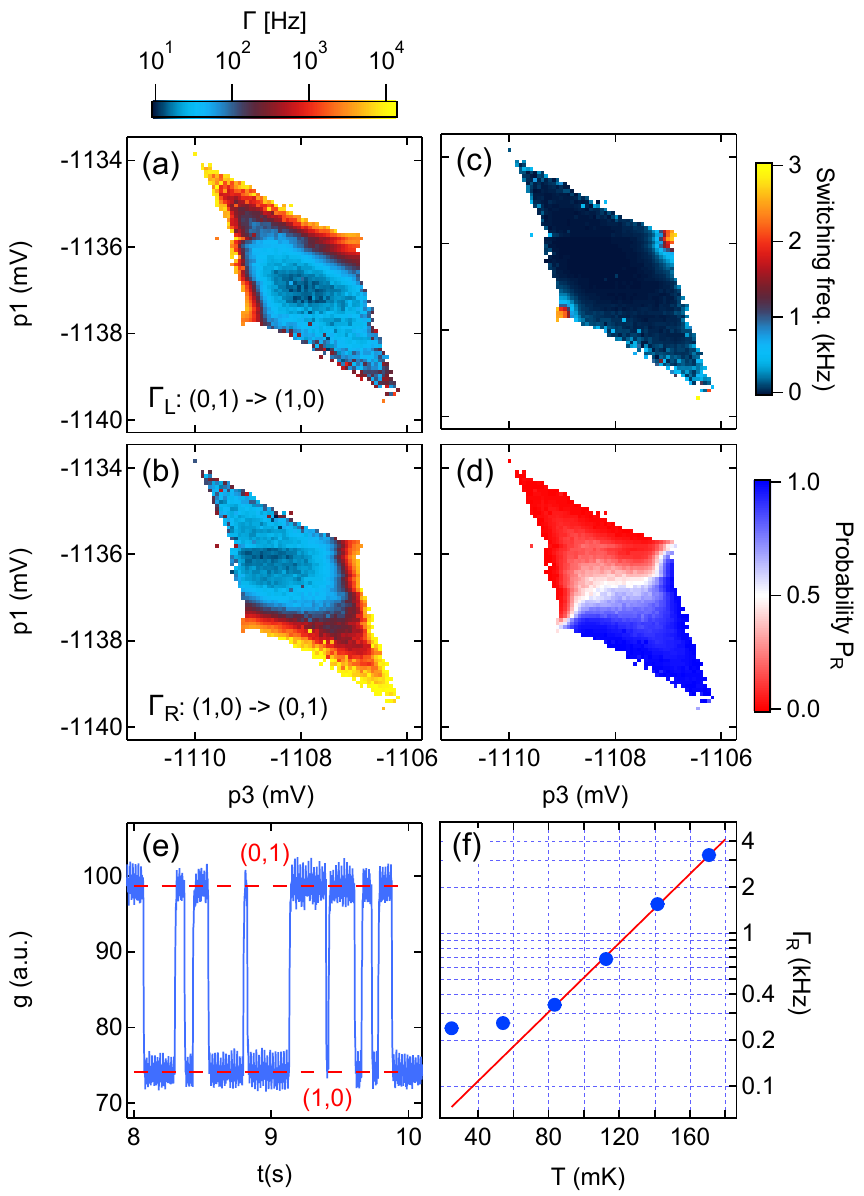}\vspace{-2mm}
\caption{(a),(b) Switching rates on a log color scale. (c) Switching frequency $f$ on a linear color scale. (d) Probability $P_{R}$ of finding an electron in the right dot. (e) Real-time trace and (f) $\Gamma_R(T)$ with exponential fit (red) to the high-$T$ data, taken at the center of the diamond [white dot in Fig.\,\ref{fig:2}(b)].}
\label{fig:3}\vspace{-4mm}
\end{figure}

The CSD drastically changes upon reduction of the interdot and reservoir tunnel rates, see Fig.~\ref{fig:2}. The zero detuning line is no longer present, transforming instead into a broad region where the DD enters a metastable charge state, repeatedly switching between the $(1,0)$ and $(0,1)$ configurations over time, see Fig.\,\ref{fig:3}(e). The switching is also visible in the standard deviation of the sensor signal, see Fig.\,\ref{fig:2}(b), and fits very well into a diamond drawn with lines following the slopes of the reservoir transitions. This suggests involvement of the reservoirs in the switching process. The measurement bandwidth is too low here to resolve the reservoir transitions in the standard deviation, thus decreasing the size of the diamond compared to a straight continuation of the reservoir lines.

To quantify the switching rates, we gather numerous events throughout the diamond and histogram the dwell times in either charge state, finding single exponential decays. The rate $\Gamma_{L}$ for switching into the left dot is given by the total number of switches $N$ into the left dot and the accumulated waiting time $t_{R}$ in the right dot, $\Gamma_L:=\Gamma_{(0,1)\rightarrow(1,0)}=N/t_R$. Similarly, $\Gamma_R$ is obtained, and $\Gamma_{L,R}$ are shown in Fig.~\ref{fig:3}(a,b). These rates are largest along the upper (a) or lower (b) edges, where a dot level approaches its reservoir level, here incidentally reaching the sensor bandwidth. Away from these edges, the rates fall off exponentially by three orders of magnitude until a small background rate is reached which varies weakly with gate voltages and is well above the smallest detectable rate. Photon assisted tunneling, residual interdot tunneling or a higher order process, possibly involving phonons \cite{KornichPRB}, could give rise to such a background rate.

The switching frequency $f=(\Gamma_L^{-1}+\Gamma_R^{-1})^{-1}$ of the complete process, i.e. from left to right dot and back, peaks close to the triple points, see Fig.\,\ref{fig:3}(c), and is low elsewhere. The probability $P_{R}$ of finding an electron in the right dot is $P_{R} = \Gamma_{L}/(\Gamma_L+\Gamma_R)$, shown in Fig.~\ref{fig:3}(d), reproducing the standard DD CSD with equal probabilities for an electron on the left or right dot at the zero detuning line. In addition, the measured temperature dependence of $\Gamma_R$ is shown in Fig.\,\ref{fig:3}(f), recorded at the center of the diamond (white dot Fig.~\ref{fig:2}(b)) for slightly different gate voltages. The rate decreases exponentially with temperature, indicating a thermally activated process, until a background rate as described before is reached, giving an upper limit to the electron temperature. The exponential decay is potentially useful for thermometry \cite{Feschchenko}, and again suggests involvement of the reservoirs with their exponential tails of the Fermi-Dirac distributions. At elevated temperatures, the switching rate exceeds the sensor bandwidth, rendering the switching diamond invisible, as observed in the experiment.

We note that the metastability shifts together with the DD triple points upon gate voltage adjustments. In addition, it disappears when introducing significant interdot tunneling, and is visible also at higher vertices, see red circles in Fig.~\ref{fig:1}(c). This confirms the DD itself as the source of the switching, rather than accidental charge traps \cite{JungAPL,PioroLadrierePRB,BuizertPRL,LiangAPL,HitachiAPL,PaladinoRMP}. Also, the switching diamond does not exhibit gate-hysteresis, ruling out latching effects. Further, the switching persists when reducing the sensor bias to 5~$\mu$V, far below the $\Delta\sim150\,\mu$eV of the short diamond axis. This rules out sensor back action, which is observed at larger biases and exhibits a pronounced bias dependence \cite{OnacPRL,TaubertPRL,GasserPRB,HarbuschPRL,GrangerNP,BraakmanAPL}, unlike the metastability seen here. Thus, this suggests an intrinsic DD process rather than an extrinsic effect as the origin for the switching.

\begin{figure}[Ht]
\vspace{0mm}\includegraphics[width=0.9\columnwidth]{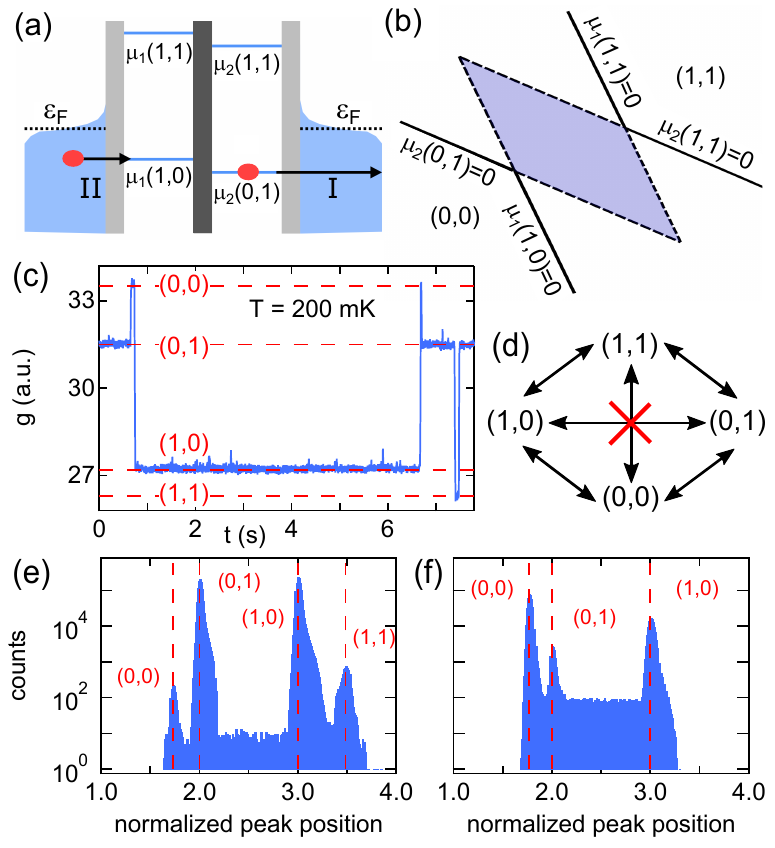}\vspace{-2mm}
\caption{(a) Illustration of the model: I) slow, thermally activated tunneling into the reservoir, followed by II) tunneling into the metastable $(1,0)$ state. (b) Both $\mu_1(1,0)$ and $\mu_2(0,1)$ but not $\mu_{1,2}(1,1)$ are below the reservoir Fermi energy within the shaded diamond of metastability. (c) Real-time sensor data for reduced reservoir coupling at 200\,mK in the center of the diamond [white dot, Fig.\,\ref{fig:2}(b)]. Switching via the intermediate states $(0,0)$ and $(1,1)$ is seen. (d) Four state Markov chain of the model. (e) Histogram of data from (c) as a function of sensor signal, normalized to the $(0,1)$ and $(1,0)$ positions. (f) For comparison, at the $(0,0)$ triple point [yellow dot, Fig.\,\ref{fig:2}(b)], the $(1,1)$ state is not seen. }
\label{fig:4}\vspace{-4mm}
\end{figure}

Based on these observations, we propose a model of thermal electron exchange. Inside the diamond, both DD one-electron levels lie below the reservoir Fermi level $\varepsilon_F=0$, see Fig.~\ref{fig:4}(a,b). From a $(0,1)$ initial state (ground state), the electron tunnels into a thermally activated hole at the same energy in the adjacent reservoir, bringing the DD into $(0,0)$. This is very slow since such a state lies in the exponentially suppressed tail of the Fermi function, setting the overall time scale for the slow switching rates. Then, an electron can either fill the initial state $(0,1)$ and restart the process, or it can end up in $(1,0)$. Both of these transitions occur quickly -- with essentially bare tunnel rates -- since the reservoir states at the corresponding dot energies are fully occupied. Given negligible interdot tunneling, the $(1,0)$ state is metastable, i.e. long lived, and is detected by the charge sensor. Alternatively, the process can go through $(1,1)$ instead of $(0,0)$ in a similar way. 

To test this model in the experiment, we note that it predicts the intermediate states $(0,0)$ and $(1,1)$ which were not seen so far (see Fig.\,\ref{fig:3}(e)). If these are occupied long enough, they are detectable with the charge sensor. Thus, we decrease the tunnel coupling to the reservoirs, increasing the time spent in the intermediate states. Further, we heat the sample to 200\,mK in order to obtain enough switches per unit time despite decreased reservoir coupling. With these changes, real-time sensor data is recorded, see Fig.\,\ref{fig:4}(c). Here, starting with $(0,1)$, the DD goes briefly to the intermediate state $(0,0)$ before switching to $(1,0)$. The DD is also sometimes seen in the $(1,1)$ intermediate state, e.g. after $t=7\,$s, here returning to where it was before the excursion to $(1,1)$. Thus, we indeed observe the intermediate states as predicted.

In the following, we provide an extension of the orthodox theory for transport in DDs \cite{vanderWiel:2002} incorporating charge fluctuations. We note that the switching rates are slow enough to validate a semi-classical description with definite occupation numbers $N_j$ of dots $j=1,2$. The charge fluctuations lead to switching between different configurations $x=(N_1,N_2)$, which we express through a master equation for the occupation probabilities $P_x$,
\begin{equation} \label{eq:master_eq}
	\partial_t P_x = \sum_{x' \neq x}
	\left[ P_{x'} \Gamma_{x'\to x} - P_x \Gamma_{x \to x'} \right],\vspace{-1mm}
\end{equation}
\noindent with $\Gamma_{x' \to x}$ the tunneling rate from configuration $x'$ to $x$. In accordance with the experiment, we maintain only $x=(1,0),(0,1),(0,0),(1,1)$ and neglect direct interdot tunneling. Hence, we keep only the rates $(1,0) \leftrightarrow \{(0,0),(1,1)\} \leftrightarrow (0,1)$ for Eq.\,(\ref{eq:master_eq}), see the Markov chain in Fig.\,\ref{fig:4}(d). By the Pauli principle, the bare tunneling rate $\Gamma_j$ between dot $j=1,2$ and its neighboring lead is weighted by the number of occupied lead states when tunneling onto the dot, $f(\mu_j(N_1,N_2))$, and by the number of unoccupied lead states when tunneling out of the dot, $1-f(\mu_j(N_1,N_2))$. Here $\mu_j(N_1,N_2)$ is the chemical potential of dot $j$ \cite{vanderWiel:2002}, $f(\varepsilon) = [1+\exp(\varepsilon/k_B T)]^{-1}$ the Fermi function (with Boltzmann constant $k_B$), and we have chosen the zero of energy at the Fermi level $\varepsilon_F=0$ of the unbiased leads. The energy dependence of $\Gamma_{1,2}$ \cite{MacLeanPRL,AmashaPRB} is very weak compared to the energy dependence of the Fermi functions relevant here. Assuming energy independent $\Gamma_{1,2}$, this leads to the following set of rates:
\begin{eqnarray}
	\Gamma_{(0,0) \to (0,1)} &=& \Gamma_2 f(\mu_2(0,1))				\label{eq1}\\
	\Gamma_{(0,1) \to (0,0)} &=& \Gamma_2 [1-f(\mu_2(0,1))]		\label{eq2}\\
	\Gamma_{(0,1) \to (1,1)} &=& \Gamma_1 f(\mu_1(1,1))				\label{eq3}\\
	\Gamma_{(1,1) \to (0,1)} &=& \Gamma_1 [1-f(\mu_1(1,1))]		\label{eq4}\,
\end{eqnarray}

\begin{figure}[Ht]
\vspace{0mm}\includegraphics[width=0.9\columnwidth]{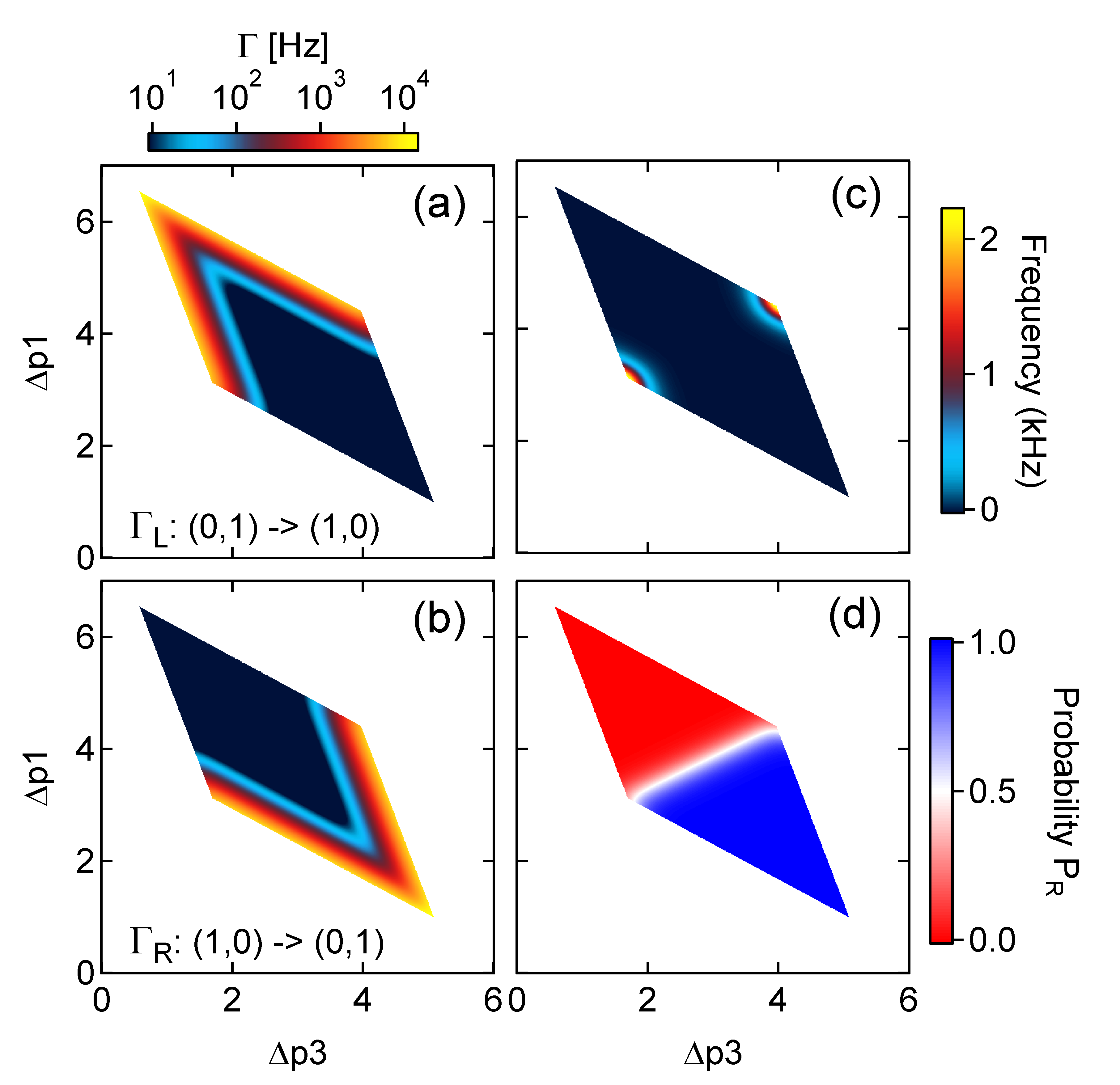}\vspace{-2mm}
\caption{Modeled quantities analogous to Fig.\,\ref{fig:3}(a-d) ($T_e=\unit[60]{mK}$, $\Gamma_{1,2}=\unit[20]{kHz}$), agreeing well with experiments. $\Gamma_{L,R}$ are shown with color scale saturated at the lowest rates seen. }
\label{fig:model}\vspace{-3mm}
\end{figure}

\noindent The remaining rates are obtained by $(0,1) \to (1,0)$ and exchanging indices $1 \leftrightarrow 2$. The stationary solution $\partial_t P_x=0$ of Eq.\,(\ref{eq:master_eq}) becomes the inversion of a $4 \times 4$ matrix and leads to the results shown in Fig.\,\ref{fig:model} (a-d), which reproduce the main features in Fig.\,\ref{fig:3}. The Fermi functions give the exponentially increasing tunnel rates when approaching the diamond boundaries. The zero detuning line exhibits no special feature in $\Gamma_{L,R}$, as also observed, since direct interdot tunneling is absent. All these features reproduce the main experimental observations.

The background rates are not captured by the model, since it does not include photon absorption, interdot tunneling and cotunneling. Interestingly, $P_R\approx 0.5$ exhibits a slight S-shape in the experiment, see Fig.\,\ref{fig:3}(d), while the model calculates a straight line, see Fig.\,\ref{fig:model}(d). We note that the S-shape approaches a straight line upon reducing sensor bias. Thus, this is a sensor back action effect \cite{Zilberberg} not included in the model. We emphasize that only this S-shape is a back action effect -- the exchange process itself is present in absence of charge sensing.

As a final test, we perform a quantitative analysis of real-time data as shown in Fig.\,\ref{fig:4}(c). We prepare histograms displaying the total time spent in each state. Close to the lower triple point, the average time spent in $(0,0)$ becomes of the same order as the time spent in $(0,1)$ and $(1,0)$, while $(1,1)$ is almost never populated \cite{Ensslin2}, as seen in Fig.\,\ref{fig:4}(f). In the center of the diamond, the DD spends most of its time in $(0,1)$ and $(1,0)$, and is rarely in $(0,0)$ or $(1,1)$, see Fig.\,\ref{fig:4}(e). Pairs of similar height peaks result when $\Gamma_1\sim\Gamma_2$. The ratio of the large to small peak height is given by a Boltzmann factor $\exp(\Delta\varepsilon/(k_B T_e))$, where $\Delta\varepsilon$ is the energy difference between DD levels and reservoir. With $T_e=200$\,mK, good agreement is found with $\Delta\varepsilon$ obtained using the lever arm extracted from fitting a Fermi function to the reservoir transitions. Thus, this again confirms the model.

In summary, we report intrinsic metastable charge state switching within diamond shaped regions in a DD. The thermally activated fluctuations involve a fast electron exchange with the leads, leading to an apparent tunneling between left and right dot when direct interdot tunneling is negligible. An extended theory explains the observations very well and predicts intermediate charge states which are observed when reducing the reservoir tunnel rates below the sensor bandwidth. We emphasize that such thermally activated exchange continues to occur when interdot tunneling is present -- i.e. in absence of metastability -- or even outside the diamond region. In these cases, it is directly detectable via intermediate states only when the sensor bandwidth exceeds the bare reservoir tunnel rates, but is invisible otherwise, though it continues to occur, possibly limiting $T_1$.

The exchange of a dot electron with a reservoir electron randomizes the DD state, thus setting an upper limit to $T_1$ for both charge and spin qubits. Note that such electron reservoir exchange appears also at other vertices, e.g. $(1,1)-(0,2)$ with singlet-triplet states, persists in a magnetic field, and is generically present in any single, double, or multiple dot coupled to a reservoir, irrespective of the host material. Taking typical values $\Delta\varepsilon= \Delta/2=75\,\mathrm{\mu eV}$, $\Gamma_{1,2}=20\,$MHz, $T_e=100\,$mK, one obtains an upper bound $T_1=\Gamma^{-1}\exp(\Delta\varepsilon/(k_B T))\sim 0.3\,\mathrm{m}$s at the center of the diamond. Shifting the occupied state towards the reservoir Fermi energy exponentially enhances the exchange process, facilitating fast gate-controlled spin initialization \cite{Shulman2012}. Decreasing $\Gamma$ linearly extends $T_1$, while decreasing temperature or increasing the energy splitting $\Delta\varepsilon$ using gate voltages does so exponentially, until the spin-phonon coupling dominates or another process such as the background rate or sensor back action limits $T_1$.

\begin{acknowledgments}
We thank H. Bluhm, D. Loss, D. Stepanenko, A. Yacoby and O. Zilberberg for valuable input and inspiring discussions and M. Br\"{u}hlmann for experimental support. This work was supported by the Swiss Nanoscience Institute (SNI), NCCR QSIT, Swiss NSF, ERC starting grant, and EU-FP7 SOLID and MICROKELVIN.
\end{acknowledgments}
\vspace{-3mm}



\end{document}